# Deep Learning Pulmonary Vessel Segmentation for Contrast-Enhanced and Non-contrast CT Images and Clinical Evaluation

**Ying Ming1, Shaoze Luo2, Longfei Zhao2, Ruijie Zhao1, Bing Li2, Qiqi Xu2, and Wei Song1***

**1 Department of Radiology, Peking Union Medical College Hospital, Chinese Academy of Medical Sciences and Peking Union Medical College, Beijing 100730, China**

**2 Research and Development Center, Canon Medical Systems (China), No.3, Xinyuan South Road, Chaoyang District, Beijing 100027, China**

## ABSTRACT

Accurate segmentation of pulmonary vessels plays a very critical role in diagnosing and assessing various lung diseases. Currently, many automated algorithms are primarily targeted at CTPA (Computed Tomography Pulmonary Angiography) types of data. However, the segmentation precision of these methods is insufficient, and support for NCCT (Non-Contrast Computed Tomography) types of data is also a requirement in some clinical scenarios. In this study, we propose a 3D image segmentation algorithm for automated pulmonary vessel segmentation from both contrast-enhanced and non-contrast CT images. In the network, we designed a Vessel Lumen Structure Optimization Module (VLSOM), which extracts the centerline (Cl) of vessels and adjusts the weights based on the positional information and adds a Cl-Dice Loss to supervise the stability of the vessels structure. We used 427 sets of high-precision annotated CT data from multiple vendors and countries to train the model and achieved Cl-DICE, Cl-Recall, and Recall values of 0.892, 0.861, 0.924 for CTPA data and 0.925, 0.903, 0.949 for NCCT data. This shows that our model has achieved good performance in both accuracy and completeness of pulmonary vessel segmentation. We finally conducted a clinical visual assessment on an independent external test dataset. The average score for accuracy and robustness, branch abundance, assistance for diagnosis and vascular continuity are 4.26, 4.17, 4.33, 3.83 respectively while the full score is 5. These results highlight the great potential of this method in clinical application.

## Introduction

Accurate and comprehensive segmentation of pulmonary vasculature is essential for diagnosing various pulmonary diseases, planning treatments, and monitoring disease progression[1,2,3]. Computed Tomography Pulmonary Angiography (CTPA) and Non-Contrast Computed Tomography (NCCT) are two widely used imaging modalities in clinical practice. While CTPA provides detailed vascular information due to the administration of contrast agents, NCCT lacks such enhancement but remains valuable for specific clinical scenarios, such as patients with contraindications to contrast media. Developing a unified framework that can effectively process both CTPA and NCCT data is therefore highly desirable[4,5]. Current research on pulmonary vessel segmentation models by deep learning focuses mostly on CTPA data, while there is less research on NCCT[6, 7]. However, NCCT is also worthy of model attention due to its easier access and lower radiation dose[8,9]. Moreover, existing works on pulmonary vessel segmentation often do not consider the integrity of the vascular lumen structure, focusing instead primarily on global statistical metrics such as DICE[10,11]. This tendency can easily lead to segmentation models being unable to effectively identify and segment small vessels or narrowed vessels adjacent to lesions, thereby impacting vessel connectivity and integrity.

In this study, we propose a deep learning model aiming to: (a) get high-precision and high-integrity pulmonary vessel segmentation, (b) support both CTPA and NCCT data, (c) exhibits robust performance for various lung diseases. A key innovation of our approach lies in the design of a Vessel Lumen Structure Optimization Module (VLSOM). This module enhances the traditional 3D U-Net architecture by extracting vessel centerlines (Cl) and applying weight-adjusted supervision for vessel lumens based on their proximity to the centerlines. By incorporating these structural cues, the model achieves high segmentation accuracy while preserving the completeness of vascular structures. In terms of ground truth (GT) annotation, we have designed an annotation scheme that allows us to simultaneously obtain high-precision GT results for both CTPA and NCCT. This enables the trained segmentation model to support automatic segmentation for both types of data. In terms of evaluation, we added Cl-DICE[17] and Cl-Recall[18] to evaluate the results, emphasizing the quantitative analysis of the accuracy and integrity of the vascular lumen structure.

The resulting pulmonary vessel segmentation algorithm offers several advantages: high precision, comprehensive vascular representation, and compatibility with both CTPA and NCCT data. These features make it a promising tool for aiding clinical diagnosis and analysis. In the subsequent sections, we will elaborate on the methodology, experimental results, and potential applications of this AI-driven model in pulmonary imaging.

Furthermore, our model demonstrates strong performance in segmenting pulmonary vessels from CTPA data, enabling precise separation of arteries and veins. More importantly, through segmentation performance transfer from CTPA to NCCT, the model successfully extends its applicability to NCCT data. This capability not only addresses the limitations of existing methods that typically focus on a single modality but also broadens the clinical utility of the proposed algorithm.

## 1. Data and Annotation

### 1.1. Internal Data information

In this study, a total of 427 cases were collected to develop and evaluate the proposed pulmonary vessel segmentation model. The dataset consisted of two main sources: 112 cases from the publicly available LIDC (Lung Image Database Consortium) dataset[12] and 315 cases from multiple clinical institutions with the approval of the institutional review boards. To ensure comprehensive support for both CTPA and NCCT data, the dataset included 213 cases of CTPA and 214 cases of NCCT cases. This balanced distribution across modalities facilitated the development of a versatile model capable of handling both contrast-enhanced and non-contrast imaging data.

The dataset was divided into two subsets for model development: 370 cases for training and 57 cases for internal testing. The training set contained 185 CTPA cases and 185 NCCT cases, while the internal testing set included 28 CTPA cases and 29 NCCT cases. This stratified partitioning strategy ensured that the model was trained and evaluated on data that reflected the diversity of clinical scenarios encountered in real-world applications.

The scanning equipment used for the mentioned experimental data included devices from Canon, GE, Siemens, and Philips, involving diseases such as pulmonary nodules, calcification, emphysema, pneumonia, pleural effusion, pulmonary mass, etc. The data sources encompass multiple hospitals from Asian countries as well as European countries. In all experimental data, the Z-direction resolution of CTPA ranges from 0.4 to 1.5 mm, the Z-direction resolution of NCCT ranges from 0.5 to 1.25 mm.

### 1.2. Internal Data Annotation

Ground truth (GT) annotations were performed by two radiologists with more than ten years of working experience to ensure high-quality supervision during model training. The annotation standard required labeling vessels as small as 1 mm in diameter, ensuring that even fine vascular structures were captured. This meticulous annotation process was critical for achieving high precision in vessel segmentation and preserving the completeness of the vascular network. In addition, the lung contours of each data were additionally annotated.

Annotating blood vessels in 3D CT images is a challenging task, which is difficult and labor-intensive for doctors to annotate directly. Therefore, we designed a data annotation scheme, and the complete ground truth (GT) generation process is shown in the **Figure1**. Given that there are already preliminary algorithms supporting automatic segmentation of CTPA data, we first used a semi-automatic segmentation method based on Hessian matrix[13] to generate an initial vascular tree (C0) on CTPA data, but it suffers from incomplete vascular branches and segmentation errors. We then used our self-developed annotation tool for doctors to carefully correct and supplement annotations (S0), ultimately obtaining a more complete and accurate annotation of the vascular branches in CTPA images (C1). To support the model's application to NCCT data, we trained a CTPA-based segmentation network using S1 and applied this network to NCCT data to obtain an initial vascular tree (N0). The above annotation process was repeated to finally achieve refined vessel annotations (N1) for NCCT data. The annotation process is shown in **Figure 1**, and the GT example of this dataset is shown in **Figure 2**.

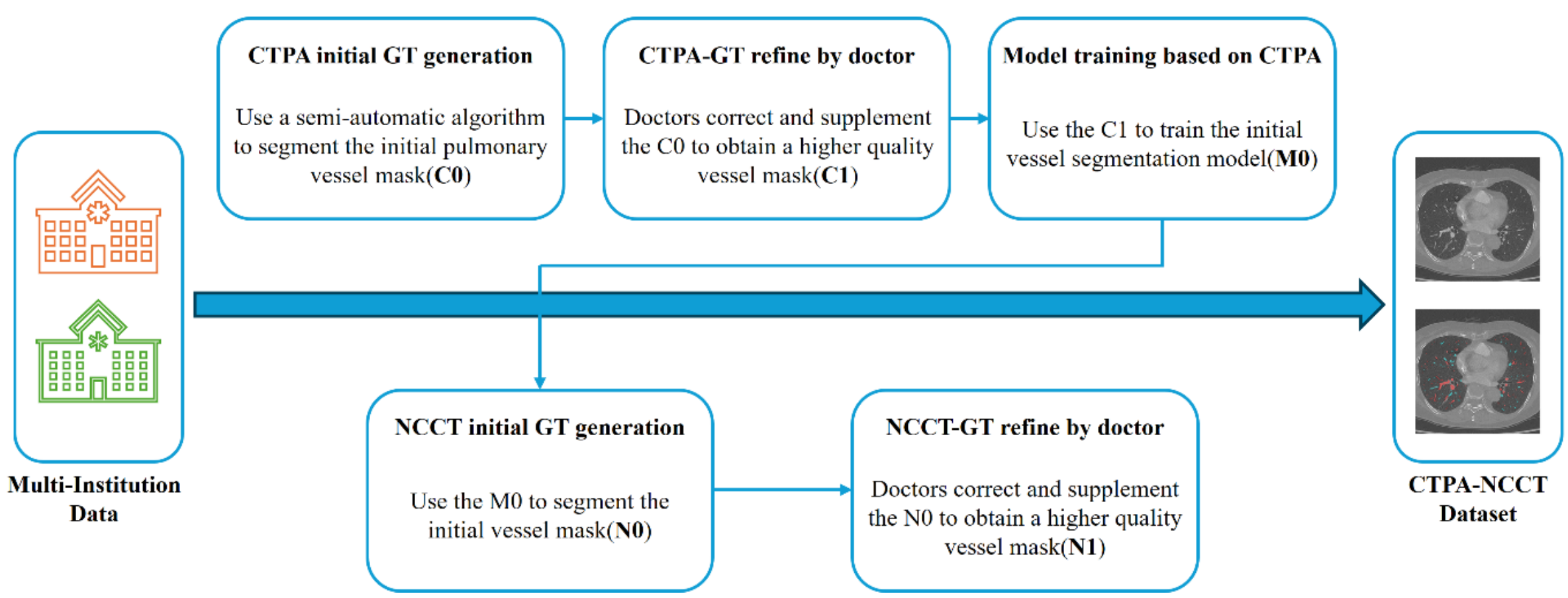

**Figure1 Iterative generation process of GT annotations of CTPA-NCCT**

In summary, the dataset used in this study was carefully curated to include a diverse range of cases from both public and private sources, with balanced representation of CTPA and NCCT data. The high-quality annotations and thoughtful data partitioning provided a strong foundation for developing a clinically relevant and versatile pulmonary vessel segmentation model.

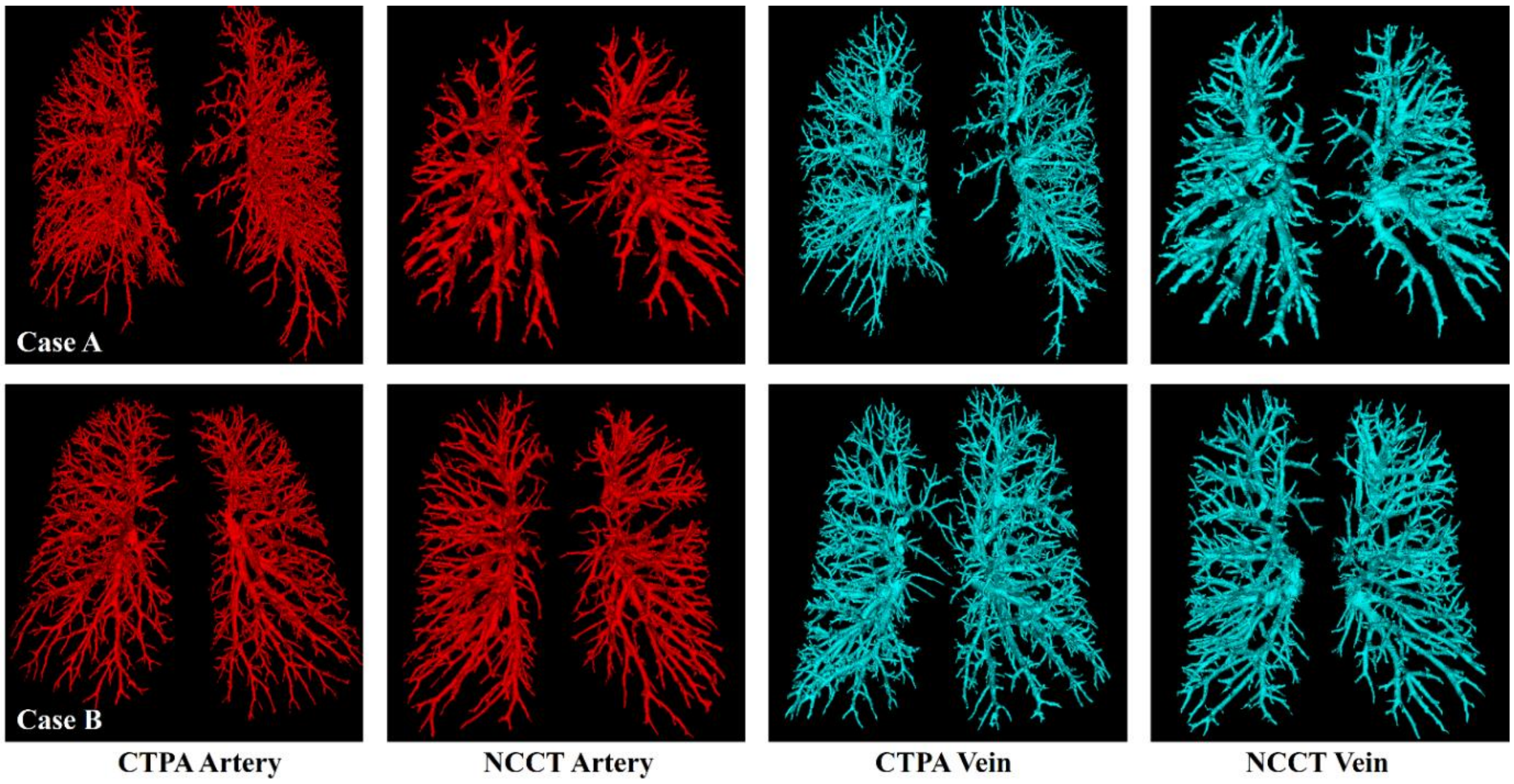

**Figure2 GT examples of experimental datasets**

## 1.3.External Evaluation Data

For the clinical visual evaluation, we collected 21 CT dataset from Peking Union Medical College Hospital from 21 patients with different lung diseases, comprising 6 males and 15 females. Among them, 6 patients underwent CTPA, including 3 cases of pulmonary vasculitis (PV), 3 cases of pulmonary embolism (PE). To demonstrate the applicability of the proposed method to NCCT, we also collected 15 cases of patients who underwent NCCT, including 4 cases of interstitial lung disease (ILD), 4 cases of chronic obstructive pulmonary disease (COPD), 3 cases of systemic lupus erythematosus (SLE) and 4 patients with ground-glass opacity (GGO).

All CTPA and NCCT scans were acquired using three different CT systems: Somatom Definition Flash and Somatom Force (Siemens Healthineers, Forchheim, Germany); Philips IQon CT (Philips Medical Systems, Best, Netherlands); Aquilion ONE GENESIS (Canon Medical Systems, Tokyo, Japan). The participants were instructed to hold their breath before the examination. During the examination, the participants were asked to assume a supine position with both arms holding their heads. The participants held their breath at the end of deep inspiration and a whole-lung scan was performed from the apex to the base of the lung. Contrast-enhanced imaging was performed by administering 40 mL of nonionic contrast agent (Iopromide, Ultravist 370, Bayer HealthCare) via the antecubital vein at 4 mL/s, followed by 30 mL saline injection at the same rate. The CT scanning parameters were as follows: tube voltage, 120 kVp; tube current, automatic tube current modulation technology; rotation time, 0.5 s; image matrix, 512×512; collimation, 64×0.6 mm (Somatom Definition Flash, Somatom Force), 64×0.625 mm (Philips IQon CT) or 80×0.5 mm (Aquilion ONE GENESIS); pitch, 1.2 (IQon CT) and 0.984 (Aquilion 64). All images were reconstructed to a 1 mm slice thickness. All images were saved in the Digital Imaging and Communications in Medicine (DICOM) format.

## 2. Evaluation method

The pulmonary vessel segmentation was evaluated on the internal test dataset using quantitative metrics. And clinical visual evaluation was conducted on the external test dataset.

### 2.1 Quantitative Evaluation

The quantitative metrics were Centerline Dice (Cl-DICE[17]), Centerline Recall (Cl-Recall[18]), Recall and DICE. As mentioned in the introduction, the DICE value can only reflect global information but cannot access the integrity and connectivity of the vessel structure. Therefore, our work mainly focuses on centerline-related indicators, such as Cl-DICE and Cl-Recall. The Higher Cl-Recall and Cl-Dice mean that more pulmonary vessel branches are correctly separated by the model, indicating a more complete and rich vascular tree segmentation result.

### 2.2 Clinical Evaluation

Four objective indicators were used to conduct the evaluation: accuracy and robustness, branch abundance, assistance for diagnosis and vascular continuity (**Table 1**)[14]. Two radiologists (with 10 and 12 years of experience in chest CT interpretation, respectively) conducted a subjective evaluation of the segmentation results. The two radiologists assessed the results using 5-point scales for scoring in ITK-SNAP software (version 3.8.0), with the manually segmented results as the gold standard. A score of 5 represented outstanding segmentation with high consistency with the gold standard, while a score of 0 indicated completely unacceptable results. The kappa test was used to analyze the inter-observer consistency of subjective clinical evaluation (Kappa value $< 0.4$: poor consistency, $0.4 \leq$ Kappa value $< 0.75$: good consistency, Kappa value $\geq 0.75$: excellent consistency).The entire evaluation process for each case took approximately 5 minutes, including the evaluation of four indicators by radiologists. The algorithm's highly accurate and detailed segmentation performance is beneficial for our study of pulmonary anatomical structures.

**Table1 Five-point scales for subjective scoring in clinical evaluation**

| Scale and Score | Description |
|---|---|
| Accuracy and robustness | |
| 0 | The segmentation result is not available or deviates substantially from the ground truth. |
| 1 | The segmentation exhibits poor reliability and yields inaccurate or inconsistent results for the principal vascular trunk. |
| 2 | Notable deficiencies exist in segmentation accuracy and robustness, failing to accurately delineate the anatomies of the pulmonary artery and vein. |
| 3 | Though generally accurate, segmentation contains appreciable errors or misclassifications between arteries and veins. |
| 4 | Segmentation proves largely accurate and robust overall, with good resulting vessel morphology, but with minor errors in a few regions |
| 5 | Segmentation results demonstrate high accuracy, with highly faithful morphology of the segmented vessels and few errors or inconsistencies. |
| Branch abundance | |
| 0 | The segmentation result only contains the vessel trunks. |
| 1 | The segmentation result only contains 2 to 4 levels of vessel branches. Here every forked branch is defined as a new branch level. |
| 2 | The segmentation result contains 4 to 5 levels of vessel branches. |
| 3 | The segmentation result contains 5 to 7 levels of vessel branches. |
| 4 | The segmentation result contains most pulmonary branches but with a few remaining undetected. |
| 5 | The result accurately identifies and segments nearly all visible branches of the pulmonary arteries and veins on CT, yielding an overall rich vascular tree morphology. |
| Assistance for diagnosis | |
| 0 | The segmentation outcomes offer negligible diagnostic utility, with the algorithm providing minimal meaningful assistance to the clinician. |
| 1 | The algorithm affords limited diagnostic aid and retains substantial room for improvement. |
| 2 | Segmentation can assist in diagnosis, yet still necessitates manual adjustments and corrections. It may facilitate surgical planning but demonstrates difficulty in aiding disease diagnosis or severity assessment. |
| 3 | Segmentation furnishes valuable diagnostic support, and with a few human adjustments, can assist surgical planning and disease diagnosis. |
| 4 | Segmentation provides a valuable asset for diagnosis and surgical planning. The clarity of the results enable clinicians to identify and analyze the structures of pulmonary arteries and veins confidently. |
| 5 | Segmentation results address most scenarios of clinical utility and can guide surgical planning and diagnosis with few requiring modifications. |
| Vascular continuity | |
| 0 | The segmentation result is not available, the principal vascular trunk is mostly intact but most vessel branches exhibit severe fragmentation, making the vascular tree morphology difficult to recognize. |
| 1 | The 2 to 4 levels and higher vessel branches are mostly intact, but the distal branches exhibit noticeable fragmentation or missing segments, resulting in an incomplete overall structure. Here every forked branch is defined as a new branch level. |
| 2 | The 4 to 5 levels and higher vessel branches are mostly intact, but distal branches fractures affect the vascular tree morphology. |
| 3 | The 5 to 7 levels and higher vessel branches are mostly intact, although some distal branches exhibit fragmentation. The overall structure remains recognizable and does not affect the main vascular pathway. |
| 4 | The overall vascular structure is well preserved, with only minor fragmentation in a few distal small vessels. The vascular tree remains largely intact. |
| 5 | Vascular continuity is well maintained, with almost no noticeable fragmentation. The overall vascular tree morphology is highly intact. |

## 3. Segmentation Method

The pulmonary artery and vein separation algorithm is designed to classify voxels into three categories: background, pulmonary arteries, and pulmonary veins. The method involves several key steps, including data preprocessing, model design, and post-processing part. Below, we detail each component of the pipeline. The overall process of this design method is shown in **Figure 3**, and the network structure (S2-1) of the model is shown in **Figure 4**.

### 3.1 Data Preprocessing

The preprocessing pipeline plays a critical role in preparing the CT images for efficient and accurate analysis by the deep learning model. The specific preprocessing steps are outlined below: (1) Lung Region Cropping: To focus on the region of interest (ROI) and reduce unnecessary background information, the CT images are first cropped based on the bounding box of their lung masks; (2) Resampling: The cropped images are then resampled to a target isotropic resolution of 0.726mm×0.726mm×0.8mm, This step ensures uniformity in voxel size across all datasets, addressing variations in the original imaging resolutions and facilitating consistent feature extraction by the model. (3) Intensity Normalization: To improve the robustness of the model to variations in image intensity, the intensity values of the resampled images are clipped to the range between the 0.5th percentile and the 99.5th percentile of the image foreground intensity values.

### 3.2 Model Design

The backbone of our model is based on the 3D U-Net architecture[15], which has demonstrated strong performance in medical image segmentation tasks due to its encoder-decoder structure with skip connections. To improve the segmentation of vascular structures, we introduce the Vessel Lumen Structure Optimization Module (VLSOM). This module extracts the centerlines of the ground truth (GT) vessels and incorporates them as additional supervisory signals during training. The three main modules include: (1) Vessel Centerline Extraction Outline: The GT vascular masks are processed to extract their centerlines. These centerlines serve as a structural cue to guide the model in identifying and preserving the fine details of vessel lumen structures; (2) Weighted Supervision: Based on clinical experience and pre-experimental results, the VLSOM applies different weights to the vessel regions based on their proximity to the centerlines. This ensures that the model pays more attention to the centerline regions during training, thereby improving the precision of vessel segmentation; (3) Loss function design:

$$Loss = CE\ Loss + DICE\ Loss + 0.5\ CL_{DICE}\ Loss$$

$$Weight\ CE\ Loss = (\omega_{class} = 3, \omega_{cl} = 15)$$

$$Weight\ DICE\ Loss = (\omega_{class} = 3, \omega_{cl} = 15, in\ \ lung\ mask)$$

$$Weight\ CL_DICE\ Loss = (\omega_{class} = 3, \omega_{cl} = 15, in\ \ lung\ mask)$$

Here, $\omega_$class represents the training weight for the non-centerline regions of the vessels, while $\omega_cl$ represents the training weight for the centerline regions of the vessels. The notation "in lung mask" indicates that this weighting logic is only effective within the lung region and no additional weighting is applied outside the lung area. The above DICE Loss[16] and Cl DICE Loss[17] are both loss functions based on the vessel contour structure.

The network was configured with 32 base channels and 17 convolutional layers, using a patch size of [192,256,96] for efficient training. It was trained for 1000 epochs with an initial learning rate of 1e-2, which decayed by 3e-5 every 50 epochs. The Adam optimizer was used to adaptively adjust the learning rates during training, ensuring stable convergence and effective parameter optimization. We used the NVIDIA Tesla A800 80G for model training and testing.

### 3.3Post-Processing

After model inference on the cropped volume, we designed a post-processing algorithm to remove outliers and adjust disconnected vessel branches. The steps of the post-processing algorithm are as follows: (1) Connected Component Analysis: Calculate the connected components for both the artery and vein domains. (2) Remove Outliers: Eliminate any connected components that lie entirely outside the lung mask. (3) Select Largest Components: Identify and select the largest connected components for both arteries and veins. (4) Handle Disconnected Artery Components: If small disconnected artery components (size < threshold voxels) are connected to the largest vein component, reclassify these small artery components as veins and update the largest vein component. (5) Handle Disconnected Vein

Components: If small disconnected vein components (size < threshold voxels) are connected to the largest artery component, reclassify these small vein components as arteries and update the largest artery component. (6) Iterative Refinement: Repeat steps 4 and 5 iteratively until no additional small components can be connected to either the artery or vein. In this experiment, the default threshold is set to 800.

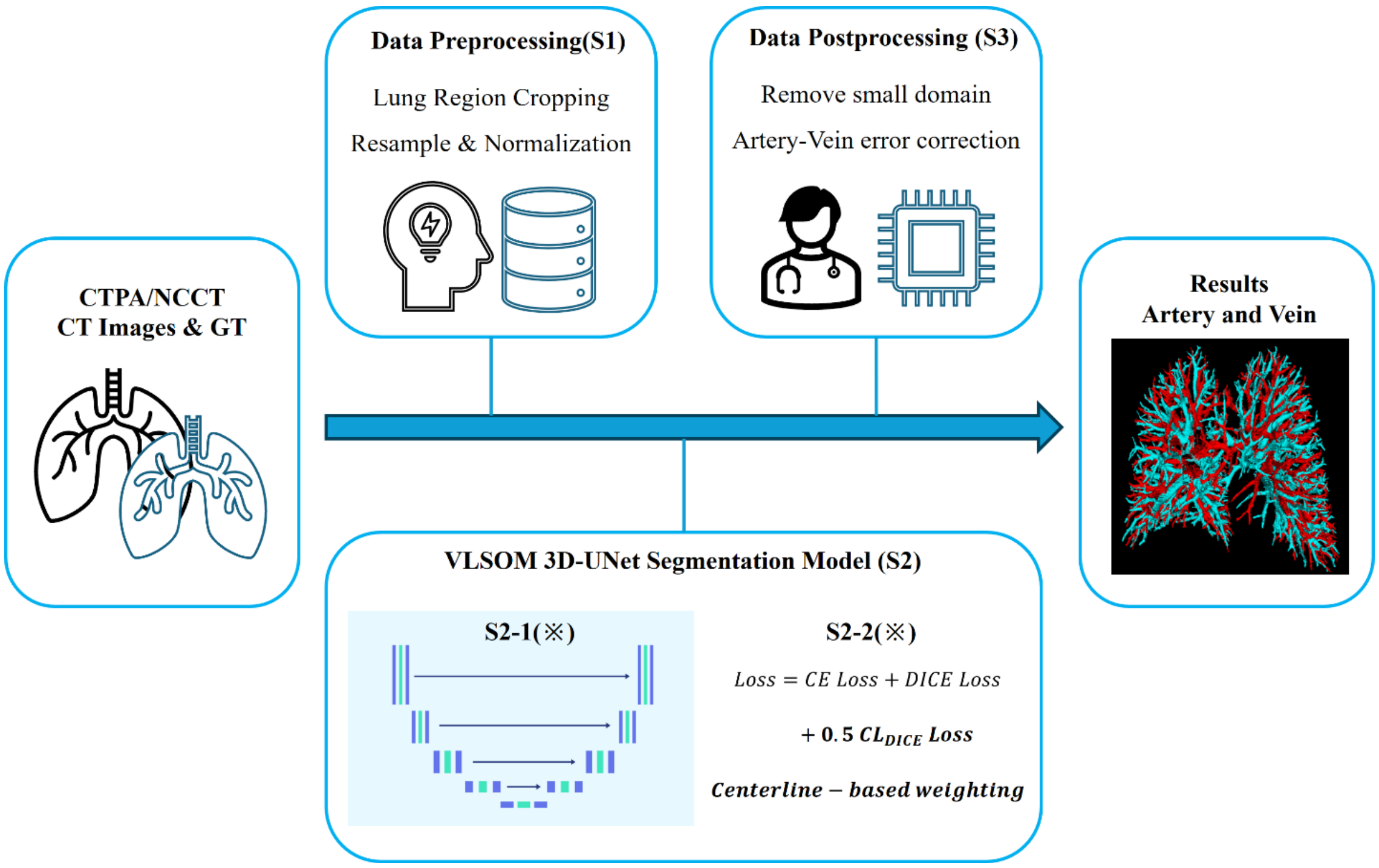

**Figure3 The main structure of the pulmonary vessel segmentation model**

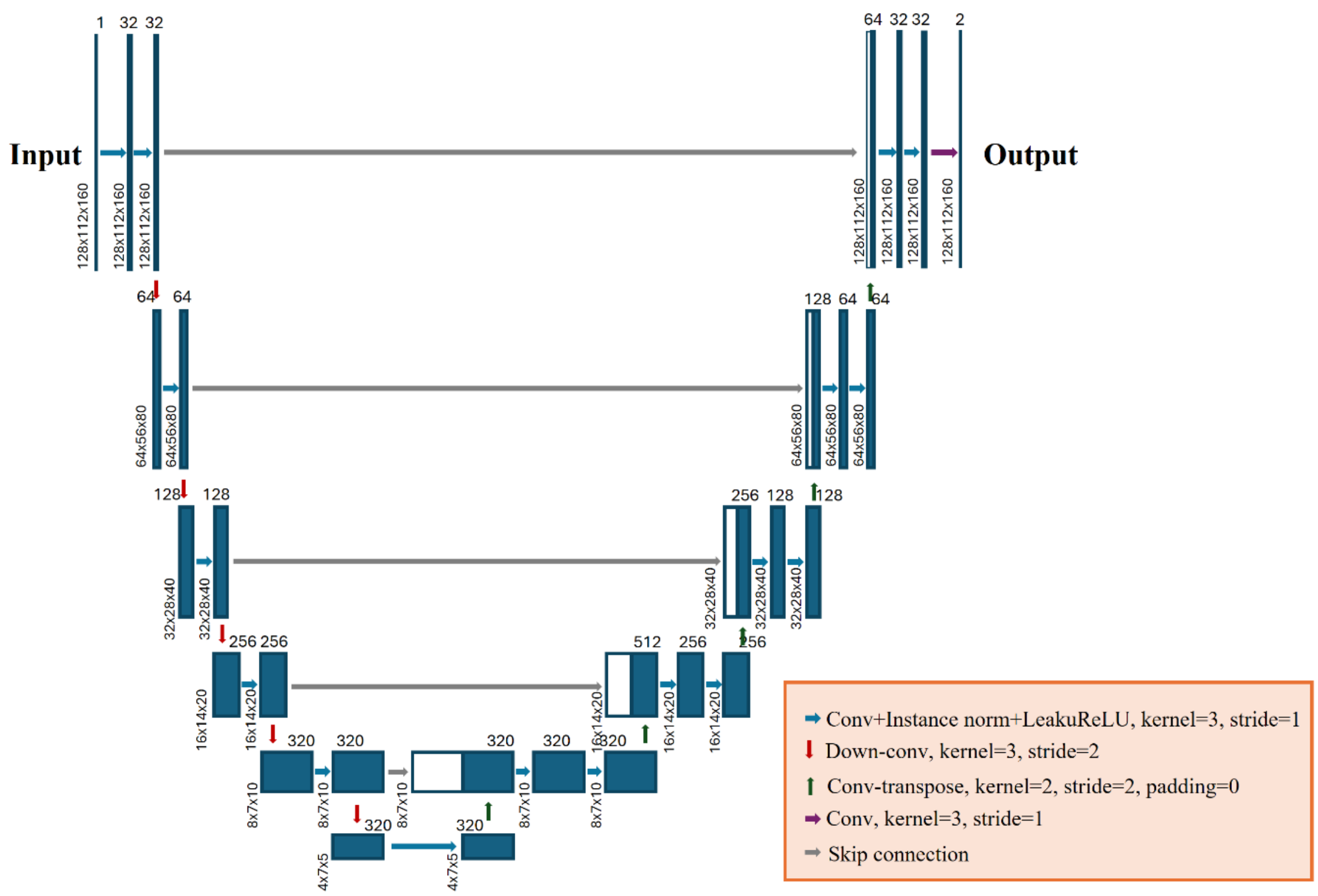

**Figure4 Specific network structure (S2-1) in the VLSOM module**

## 4. Results

During the model testing phase, we conducted evaluations using 57 datasets, including both CTPA (Computed Tomography Pulmonary Angiography) and NCCT (Non-Contrast Computed Tomography) data. The datasets were collected from multiple hospitals in China, Japan, Europe, and the Americas. The data types covered a range of pulmonary conditions, such as pulmonary nodules, emphysema, and lung masses, among others. Additionally, the data distribution included diverse ethnicities and age groups, demonstrating good generalizability. Compared with the state of the art (SOTA) method nnUNet-v2[15], our model demonstrates superior performance in pulmonary vessel segmentation for both CTPA and NCCT data. The results are shown in **Table 2**.

**Table2 Pulmonary vessel segmentation results of CTPA and NCCT data**

| CTPA | Artery Cl-DICE | Vein Cl-DICE | Artery Cl-Recall | Vein Cl-Recall | Artery Recall | Vein Recall | Artery DICE | Vein DICE |
|---|---|---|---|---|---|---|---|---|
| nnUNet-v2 | 0.816 | 0.849 | 0.710 | 0.754 | 0.820 | 0.867 | 0.863 | **0.891** |
| Our Method | **0.874** | **0.911** | **0.819** | **0.903** | **0.896** | **0.952** | **0.866** | 0.849 |
| NCCT | Artery Cl-DICE | Vein Cl-DICE | Artery Cl-Recall | Vein Cl-Recall | Artery Recall | Vein Recall | Artery DICE | Vein DICE |
| nnUNet-v2 | 0.877 | 0.895 | 0.798 | 0.823 | 0.882 | 0.873 | 0.901 | **0.882** |
| Our Method | **0.916** | **0.935** | **0.879** | **0.928** | **0.935** | **0.963** | **0.906** | 0.871 |

As shown in the table 2, the method proposed in this paper has achieved a relatively significant improvement over the nnUNet-v2 in terms of the Cl-Recall, Cl-Dice and Recall metrics. By averaging the results of arteries and veins, our experimental model achieved Cl-DICE, Cl-Recall and Recall values of 0.892, 0.861, 0.924 for CTPA and 0.925, 0.903, 0.949 for NCCT data. While State of the Art (nnUNet-v2) were 0.832, 0.732, 0.843 for CTPA and 0.886, 0.811, 0.878 for NCCT data. This indicates that the model in this paper can segment more complete and intricate vascular tree branches. However, when using this method, it may lead to coarser segmentation results, which in turn caused a certain decrease in the DICE metric. Nevertheless, during clinical evaluations, this slight thickening of the vessels did not significantly affect the identification and analysis of the lumen structure. From both the global and local levels, we documented the actual segmentation results of the proposed model on the test data. It can be observed that our method can obtain more complete vascular branches. The overall comparison of the segmentation results is shown in **Figure5** and **Figure6**, and the comparison of local vascular structures is shown in **Figure7** and **Figure8.** In the following illustrations, red labels represent pulmonary arteries and blue labels represent pulmonary veins.

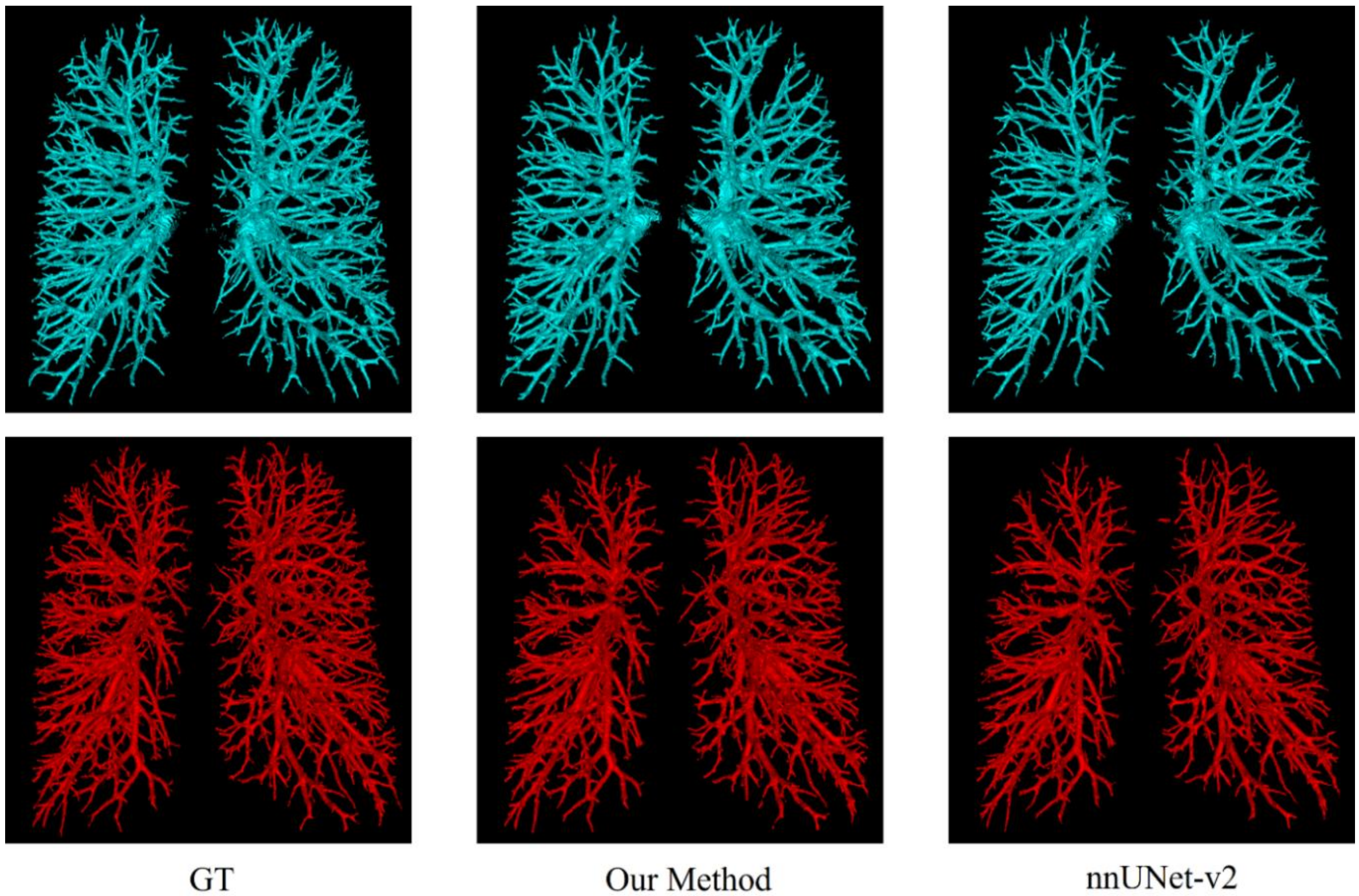

**Figure5 Example of global segmentation results of CTPA data**

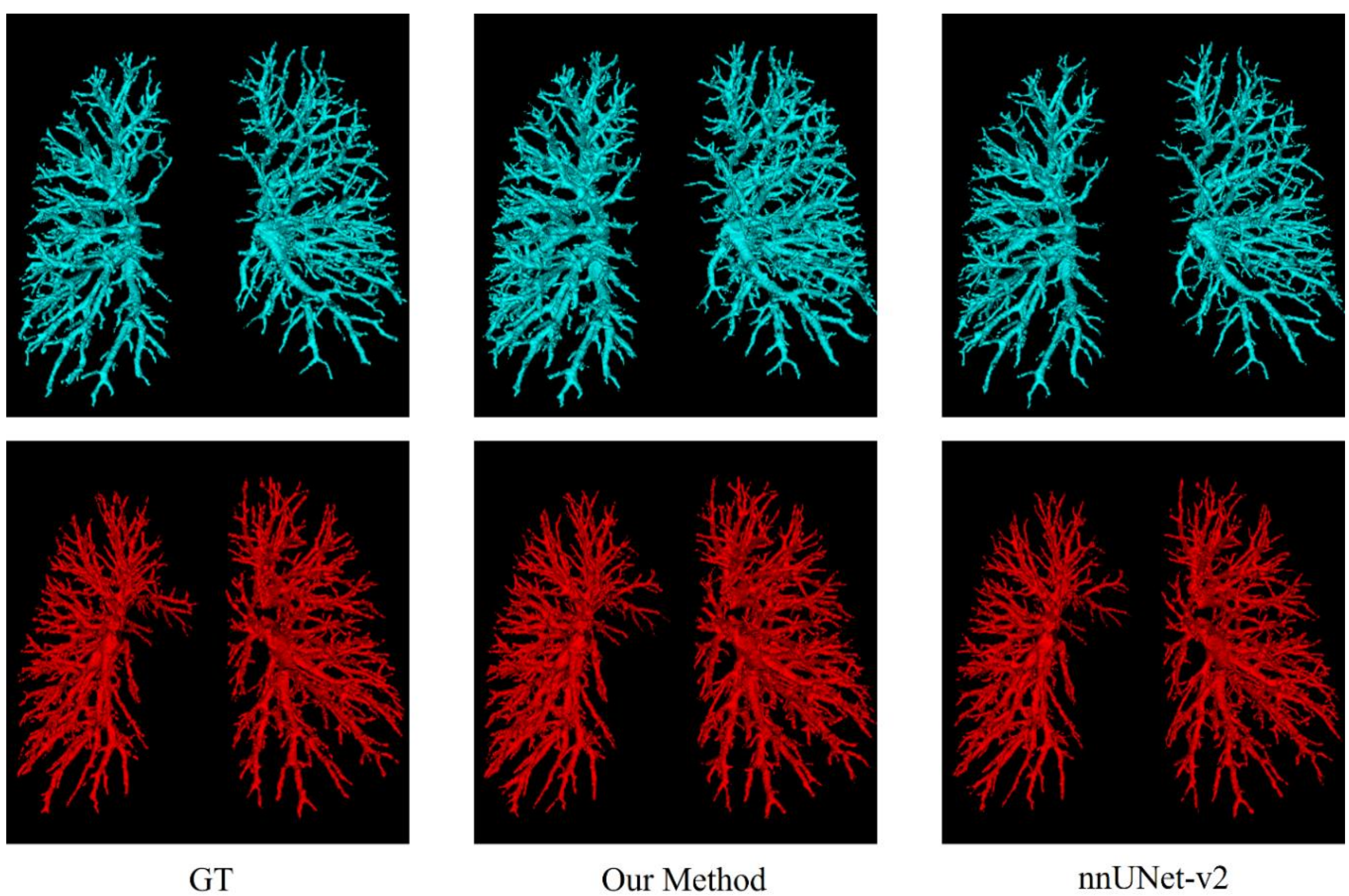

**Figure6 Example of global segmentation results of NCCT data**

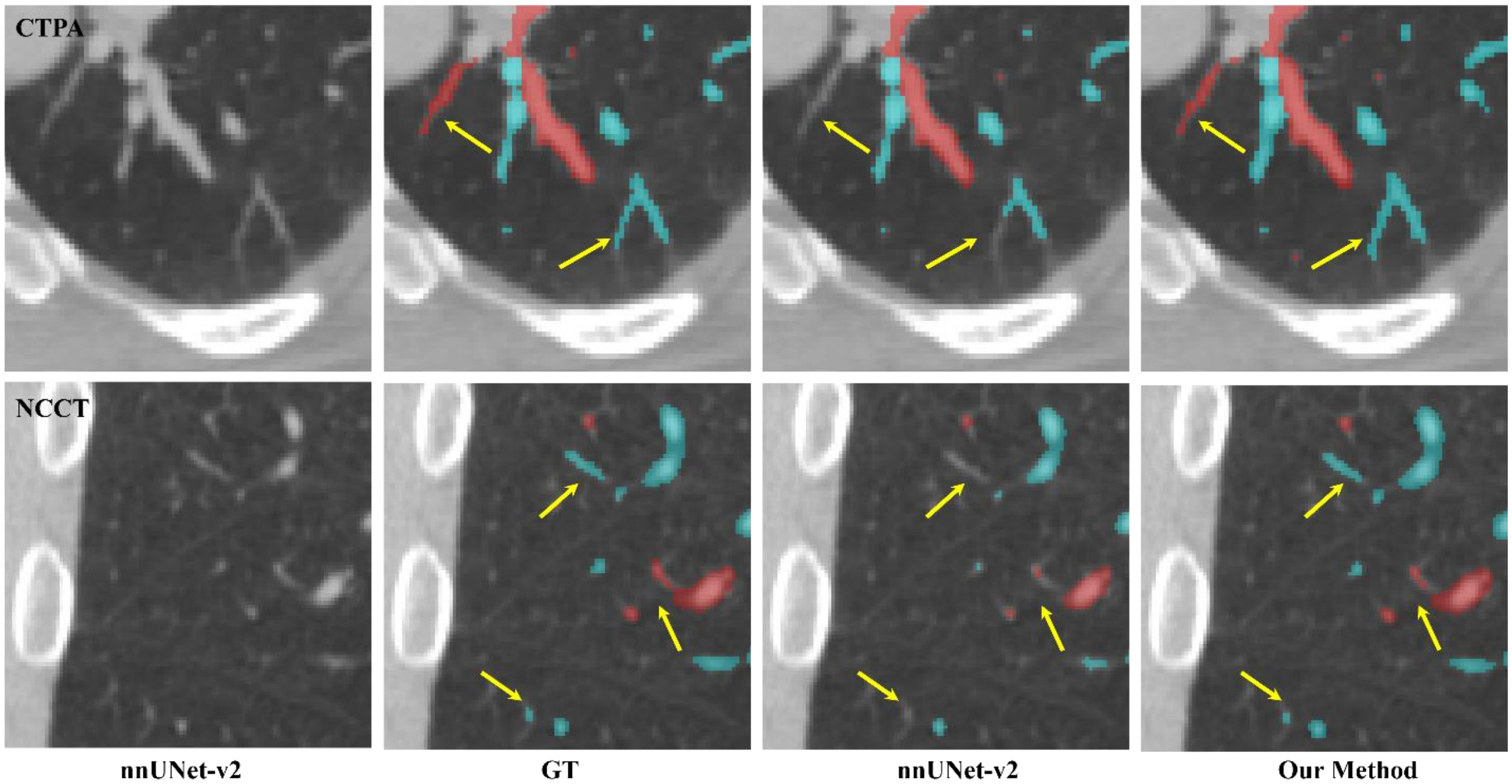

**Figure7 Example of local segmentation results for pulmonary vessels(2D)**

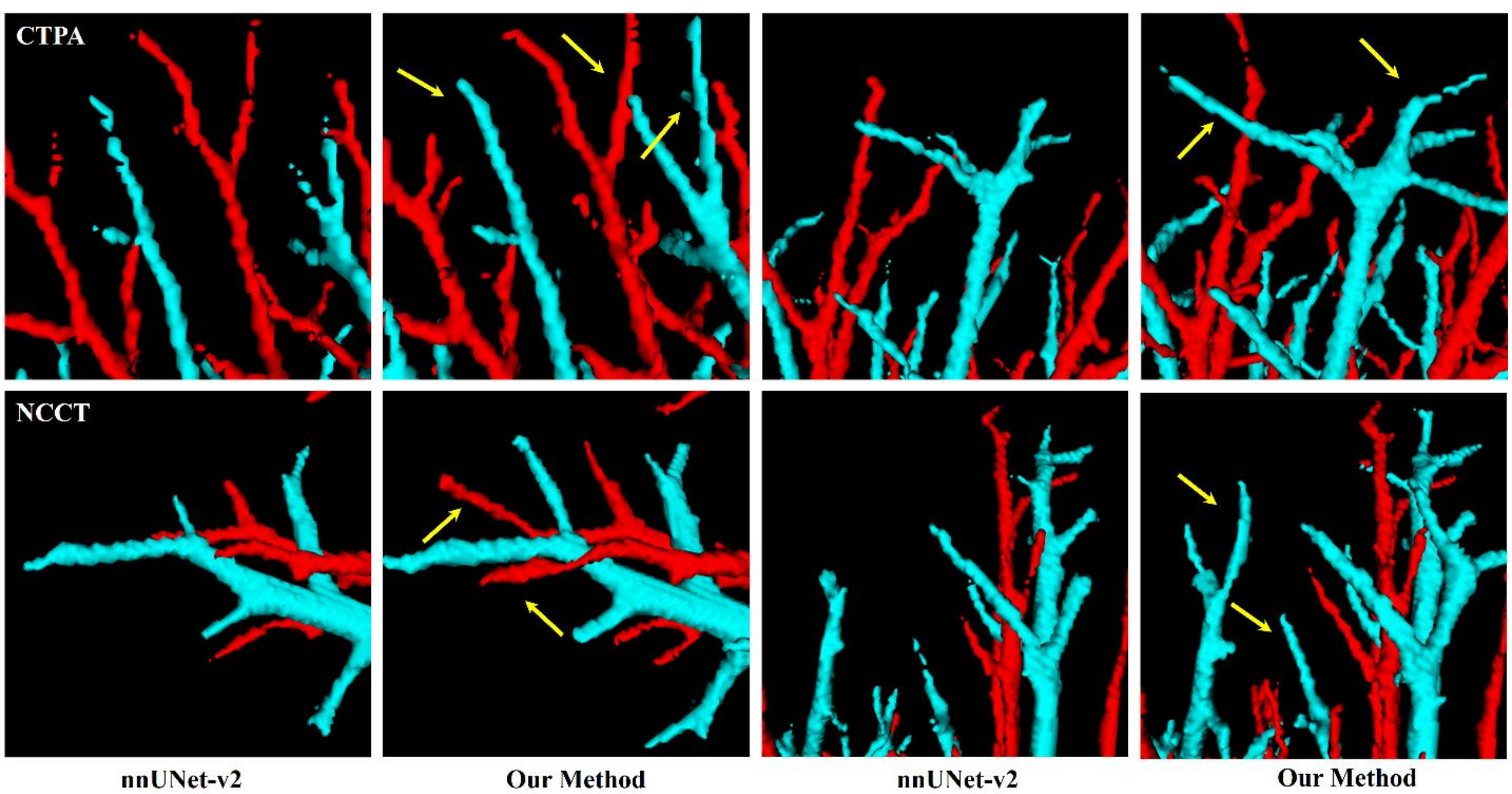

**Figure8 Example of local segmentation results for pulmonary vessels(3D)**

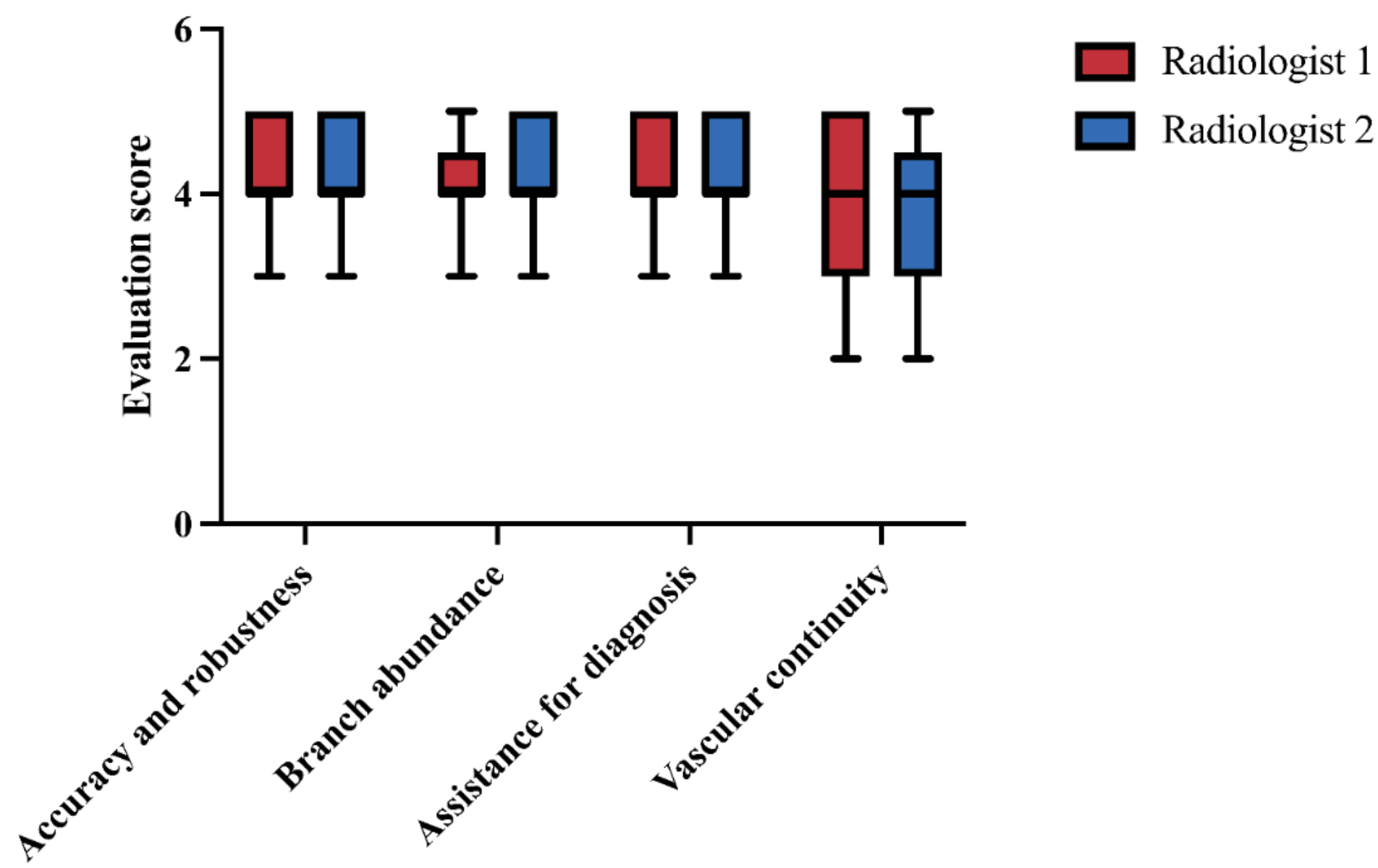

**Figure9 Clinical evaluation of four objective indicators**

We also conducted a clinical visual assessment on an independent external test dataset. The average score for accuracy and robustness, branch abundance, assistance for diagnosis and vascular continuity are 4.26, 4.17, 4.33, 3.83 respectively while the full score is 5. The consistency of clinical evaluation by two radiologists showed good consistency across the four indicators (Kappa value of 0.916, 0.744, 0.674 and 0.796, respectively). The clinical evaluation's statistical results are displayed in **Figure 9**. The scoring results indicate that our model can segment more accurate and complete vascular trees, thereby providing support for clinical diagnosis.

To compare the advantages of our approach to some existing works, we conducted vascular segmentation experiments on the Parse2022[19] dataset, and an open-source dataset provided by HiPaS model [10]. We compared our results with the ground truth (GT) provided by these works. The advantages of our method were analyzed using both visual comparisons and quantitative metrics such as Cl-Recall. We selected 50 cases from each of the two datasets for evaluation. The Parse2022 dataset only has arterial GT, so we only performed arterial segmentation on this data. The comparison between the model segmentation results and the visual results of the dataset GT is shown in **Figure 10**. And The Cl-Recall for Parse2022 and HipaS dataset are 0.973 and 0.957 respectively. From the visual and numerical results, we can see that this method can segment more complete pulmonary vessel segmentation results.

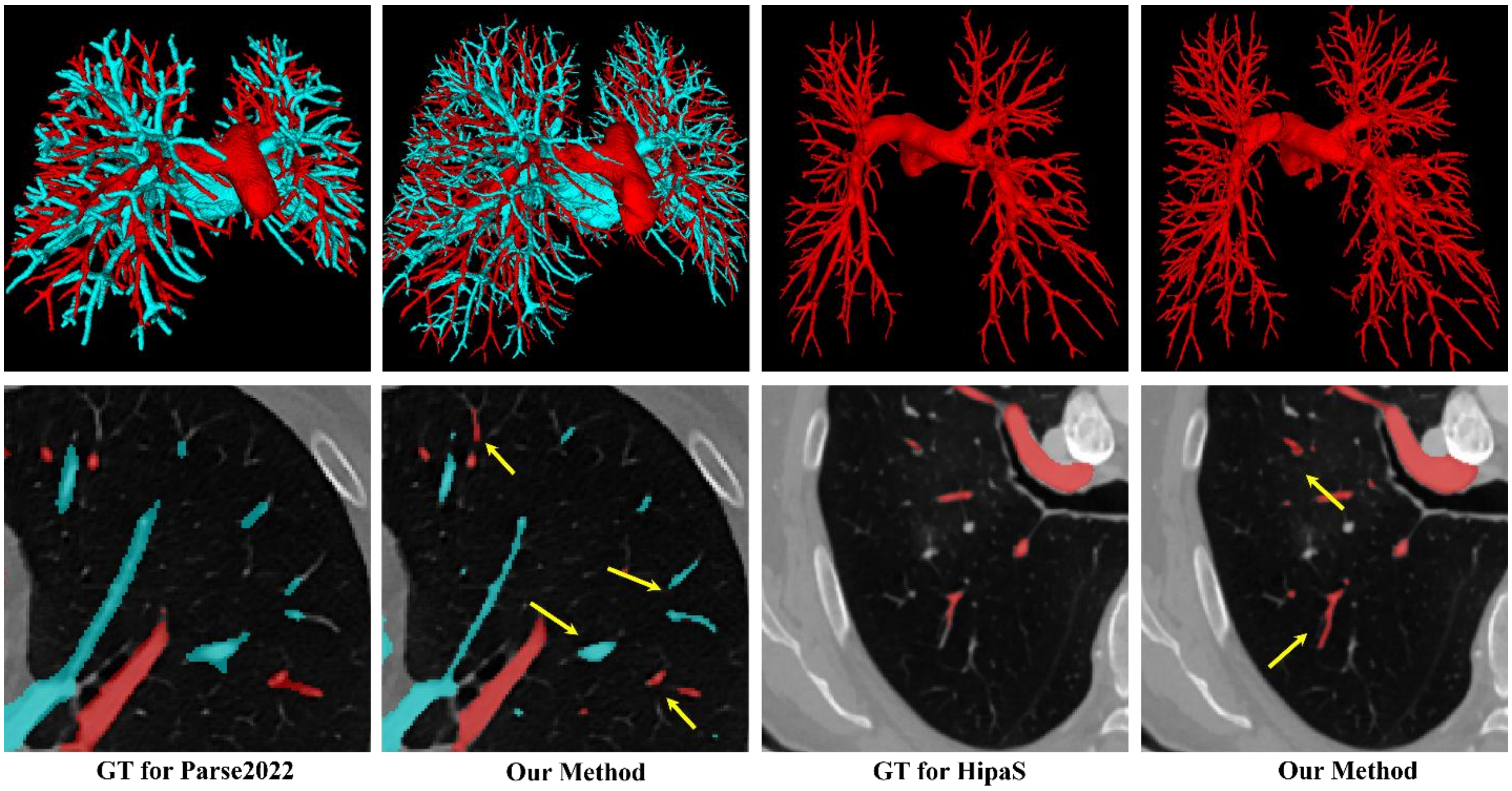

**Figure10 Visual comparison of our method and public dataset**

## 5. Discussion

This study introduced a 3D UNet segmentation network with VLSOM module to enhance pulmonary vessel segmentation performance. By leveraging combined learning from 213 CTPA and 214 NCCT data, the proposed method generated a single model which achieved high-accuracy and high-integrity automatic vascular segmentation on both 27 CTPA and 29 non-contrast CT images with Cl-Recall, Cl-DICE and Recall values of 0.879, 0.909, 0.934 for CTPA and 0.928, 0.936, 0.955 for NCCT respectively, significantly outperforming SOTA nnU-Net based model trained on the same training dataset. Furthermore, we proposed four visual assessment metrics for subjective scoring in clinical evaluation and the model demonstrated good performance on 21 independent external evaluation CT images from 21 patients with six different lung diseases (PV, PE, ILD, COPD, SLE and GGO), indicating potential for clinical pulmonary vessel analysis and disease diagnosis.

However, the current algorithm still has limitations. The remaining challenges for pulmonary vessel segmentation model include insufficient segmentation accuracy for extreme small vessels (those with diameters less than 0.8 mm) and partial errors in artery-vein classification, particularly in complex or low-contrast regions. These issues may arise due to the difficulty of capturing fine details in medical images and the complexity of pulmonary vasculature.

Future improvements could focus on incorporating advanced techniques such as multi-scale feature extraction or attention mechanisms to better detect small vessels. Additionally, expanding the training dataset and integrating anatomical priors could help reduce classification errors and improve robustness. In terms of clinical evaluation, although six different lung diseases were evaluated, the number of cases were still a little bit small due to the time cost limitation for visual evaluation. Next, we shall collect a much larger size of dataset from multiple clinical sites and validate generalization of the model performance. And further clinical studies should research the clinical values for the quantitative parameters derived from the segmented vessels such as vessel length, vessel volume, and tortuosity etc., which are potential imaging biomarkers for lung disease diagnosis and clinical decision making. Despite these limitations, the proposed method represents a significant step forward in pulmonary vessel segmentation and demonstrates promising potential for clinical applications. Further research is needed to validate its value in clinical practice.